\documentclass[showpacs,preprintnumbers,amsmath,amssymb,showkeys]{revtex4}
\def\texpsfig#1#2#3{\vbox{\kern #3\hbox{\includegraphics{#1}\kern #2}}\typeout{(#1)}}

\input epsf   

\usepackage{graphicx}
\usepackage{dcolumn}
\usepackage{bm}
\begin{document}

\title{Sine-Gordon Effective Potential beyond Gaussian
Approximation}

\author{Wen-Fa Lu (1,2,4)}
   \email{wenfalu@online.sh.cn}
\author{Chul Koo Kim (1,4)}
\author{Kyun Nahm (3)}
\affiliation{(1) \ Institute of Physics and Applied Physics,
Yonsei University, Seoul 120-749, Korea}
\affiliation{(2) \
Department of Physics, Shanghai Jiao Tong University, Shanghai
200030, the
  People's Republic of China \ (permanent
address)}
\affiliation{(3) \ Department of Physics, Yonsei
University, Wonju 220-710, Korea}
\affiliation{(4) \  Center for
Strongly Correlated Materials Research, Seoul National University,
Seoul 151-742,Korea}
\date{\today}

\begin{abstract}
Combining an optimized expansion scheme in the spirit of the
background field method with the Coleman's normal-ordering
renormalization prescription, we calculate the effective potential
of sine-Gordon field theory beyond the Gaussian approximation. The
first-order result is just the sine-Gordon Gaussian effective
potential (GEP). For the range of the coupling $\beta^2\leq
3.4\pi$ (an approximate value), a calculation with Mathematica
indicates that the result up to the second order is finite without
any further renormalization procedure and tends to improve the GEP
more substantially while $\beta^2$ increases from zero.
\end{abstract}

 \maketitle

\section{Introduction}
\label{1}

The effective potential (EP) of sine-Gordon (sG) field theory is
useful and important for studying sG field theory itself and its
equivalent models in quantum field theory and condensed matter
physics. It has been calculated with the one-loop (1L) and
Gaussian approximations, yielding the 1L and Gaussian EPs,
respectively. In 1980s, the sG 1L EP or plus its thermal
correction was used for calculating quantum or thermal correction
to the masses of the sG soliton or breathers \cite{1,2}. Analyzing
the sG 1L EP with thermal or finite-size effect gave the
information on the vacuum structure of the sG system under the
finite-temperature or -size condition \cite{3,4} in 1990s. Very
recently, the sG 1L EP was used to investigate the phase structure
of a compact U(1) gauge theory \cite{5}. The sG Gaussian EP
without or with thermal effect has been analyzed to discuss vacuum
stability and structure of the sG system \cite{6,7}. Moreover, in
1990s, the sG Gaussian EP was employed to discuss the sG field
theory with a finite cutoff \cite{8} and obtain phase diagrams of
a two-dimentional neutral classical Coulomb gas \cite{9}.
Ref.~\cite{5} also mentioned that the sG Gaussian EP can produce
interesting results on the phase structure for the above-mentioned
compact U(1) gauge theory. Besides, based on the GEP method, the
two-particle excited states can be constructed to study bound
states and scattering states in sG field theory
\cite{10}\cite{8}(1994).

In the weak coupling limit, the sG Gaussian EP can be reduced to
the sG 1L EP, and the sG Gaussian EP provides more information
than the sG 1L EP. However, no work estimated the approximation
accuracy of the sG Gaussian EP. Furthermore, the sG 1L EP is
ill-defined for the domain where the secondary derivative of the
classical potential ${\frac {\partial^2 V(\phi)}{\partial
\phi^2}}$ is negative \cite{1,2,4,5} and the curves of the sG
Gaussian EP is not smooth but sharp when ${\frac {\partial^2
V(\phi)}{\partial \phi^2}}$ vanishes \cite{6,7}. Although there
has been a long history in studying the sG system and some
interesting as well as important quantities of it have been found
or established rigorously \cite{11}, the exact EP of sG field
theory has not been obtained yet. Under such a circumstance, a
better approximated EP of sG field theory beyond the Gaussian
approximation will be worth considering and may give a substantial
correction to the sG Gaussian EP. Recently, the calculation of the
EP beyond the Gaussian approximation has attracted much interest
\cite{12,13} and the $\lambda\phi^4$ model as well as some simple
fermion models were considered \cite{12,13,14}. Very recently, we
also considered a class of scalar field theories with the
optimized Rayleigh-Schr\"{o}dinger expansion \cite{14}. Therefore,
it is interesting to add an exponential interaction model, for
example the sG field theory, to this list. This paper will report
such a work. Using an optimized expansion scheme in the spirit of
the background field method, which was developed by Stancu and
Stevenson \cite{13}, together with the Coleman's normal-ordering
renormalization prescription \cite{15,16}, we calculate the EP of
sG field theory up to the second order. No explicit divergences
exist in the resultant expression and the first-order result is
just the sG Gaussian EP. A further numerical calculation with
Mathematica indicates that the result up to the second order is
finite for the case of the coupling $\beta^2\leq 3.4\pi$ and its
correction to the Gaussian EP becomes more substantial while
$\beta^2$ increases from zero.

Next section, the optimized expansion scheme in Ref.~\cite{13} and
the Coleman's normal-ordering prescription in the functional
integral formalism \cite{16} will be introduced into sG field
theory to calculate its EP.  We will calculate the approximated sG
EP up to the second order in Sect. III and have a simple
discussion on its features in Sect. IV. A brief conclusion will be
made in Sect. V.

\section{Optimized Expansion of sG EP}
\label{2}

For the sG field theory, we use the following Lagrangian
\begin{equation}
{\cal L}_x= {\frac {1}{2}}\partial_\mu \phi_x \partial^\mu \phi_x -{\frac {m^2}{\beta^2}}[1-\cos(\beta\phi_x)] \;,
\end{equation}
where the subscript $x=({\bf x},t)$ represents the coordinates in
a $(1+1)$-dimensional Minkowski space, $\partial_\mu$ and
$\partial^\mu$ are the corresponding covariant derivatives,
$\phi_x$ the scalar field at $x$, $m$ a mass parameter and the
dimensionless $\beta$ the coupling parameter. It is always viable
to have $\beta\ge 0$ without loss of generality. Obviously, the
classical potential $V(\phi_x)={\frac
{m^2}{\beta^2}}[1-\cos(\beta\phi_x)]$ is invariant under the
transform $\phi\to \phi+{\frac {2\pi n}{\beta}}$ with any integer
$n$, and so the classical vacua are infinitely degenerate.

There exist several ways of defining the effective potential, and we will adopt the functional integral formalism
and start from the generating functional for Green's functions \cite{17}
\begin{equation}
Z_M [J]=\int {\cal D}\phi \exp\{i\int d{\bf x}dt[{\cal L}_x+J_x\phi_x]\}  \;,
\end{equation}
where $J_x$ is an external source at $x$, ${\cal D}\phi$ the
functional measure and the subscript $M$ implies Minkowski space.
In the functional integral of Eq.(2), the integrand is
oscillatory. One way of calculating $Z_M [J]$ is to transform it
into a generating functional in Euclidean space
\begin{equation}
Z[J]=\int {\cal D}\phi \exp\{-\int d^2 r[
       {\frac {1}{2}}\nabla_r \phi_r
        \nabla_r \phi_r + V(\phi_r) -J_r\phi_r]\}
\end{equation}
with a time continuation $t \to -i \tau$ \cite{17}. Here, $r=({\bf
x},\tau)$ and $\nabla_r$ is the gradient with respect to $r$ in
two-dimensional Euclidean space. Taking $W[J]=\ln(Z[J])$, which is
the generating functional for the connected Green's functions, one
can get the vacuum expectation value of the field $\phi_r$ in the
presence of $J_r$
\begin{equation}
\varphi_r={\frac {\delta W[J]}{\delta J_r}} \;.
\end{equation}
Thus, the effective potential in Euclidean space can be defined as
\begin{equation}
{\cal V}(\varphi)=-{\frac {W[J]-\int d^2 r J_r\varphi_r}{\int d^2 r}}
          \bigg|_{\varphi_r=\varphi}         \;,
\end{equation}
where $\varphi$ is independent of the coordinate $r$. In
principle, returning to Minkowski space from Eq.(5), one can get
the effective potential of sG field theory. Nevertheless, it is
not necessary to continue Eq.(5) back to Minkowski space. In fact,
the proper functions in Minkowski space coincide with those in
Euclidean space \cite{18}, and discussions on the $\lambda\phi^4$
model in Refs.~\cite{13,14,16} have suggested and supported this
point.

For any $(1+1)$-dimensional scalar field theory without derivative interactions, the Coleman's normal-ordering
prescription \cite{15} is useful for renormalizing the theory. Hence, we introduce the prescription according to
Ref.\cite{16} and rewrite $Z[J]$ as
\begin{eqnarray}
Z[J]&=&\exp\{\int d^2 r [{\frac {1}{2}}I_{(0)}({\cal M}^2) -{\frac {1}{2}}{\cal M}^2 I_{(1)}({\cal M}^2)]\}
\nonumber \\ && \times \int {\cal D}\phi
       \exp\{-\int d^2 r[ {\frac {1}{2}} \phi_r(-\nabla_r^2)\phi_r-J_r\phi_r + {\cal N}_{\cal M}[V(\phi_r)]]\}
\end{eqnarray}
where,
\begin{eqnarray*}
I_{(n)}(Q^2)\equiv\left \{
    \begin{array}{ll}
\int {\frac {d^2 p}{(2\pi)^2}}
{\frac {1}{(p^2+Q^2)^n}} \;, & \ \ \ \ \ for \ \ n\not=0  \\
 \int {\frac {d^2 p}{(2\pi)^2}} \ln(p^2+Q^2)  \;, & \ \ \ \ \
                          for \ \  n=0
    \end{array} \right.
\end{eqnarray*},
the symbol ${\cal N}_{\cal M}[\cdots]$ means the normal-ordering
form with respect to ${\cal M}$ and ${\cal N}_{\cal
M}[V(\phi_r)]={\frac
{m^2}{\beta^2}}[1-\cos(\beta\phi_r)\exp\{{\frac
{\beta^2}{2}}I_{(1)}({\cal M}^2)\}]$.

Now the optimized expansion scheme in Ref.[13] can be applied to
sG field theory. One can start it from a modification of Eq.(6)
with the following steps. First, a parameter $\mu$ is introduced
by adding a vanishing term $\int d^2 r{\frac
{1}{2}}\phi_r(\mu^2-\mu^2)\phi_r$ into the exponent of the
functional integral in Eq.(6). Then, make a shift $\phi_r \to
\phi_r+\Phi$ with $\Phi$ a constant background field. Finally,
inserting an expansion factor $\delta$, one can modify $Z[J]$ as
\begin{eqnarray}
&Z[J;&\Phi,\delta]=\exp\{\int d^2 r [{\frac {1}{2}}I_{(0)}({\cal
M}^2) -{\frac {1}{2}}{\cal M}^2 I_{(1)}({\cal M}^2)-{\frac
{m^2}{\beta^2}}+J_r\Phi]\} \nonumber
\\ && \times \int {\cal D}\phi
       \exp\{-\int d^2 r[{\frac {1}{2}} \phi_r(-\nabla_r^2+\mu^2)\phi_r-J_r\phi_r]\} \exp\{-\delta\int d^2 r {\cal
       H}_I(\phi,\Phi,\mu)\}
\end{eqnarray}
with
\begin{equation} {\cal H}_I(\phi_r,\Phi,\mu)=-{\frac {1}{2}}\mu^2\phi_r^2-{\frac
{m^2}{\beta^2}}\cos(\beta(\phi_r+\Phi))\exp\{{\frac {\beta^2}{2}}I_{(1)}({\cal M}^2)\} \;.
\end{equation}
Note that here $\delta$ is a formal parameter and will be
extrapolated to one at the final step. Further, $Z[J;\Phi,\delta]$
can be rewritten as
\begin{eqnarray}
Z[J;\Phi,\delta]&=&\exp\{-\int d^2 r [{\frac {1}{2}}
I_{(0)}(\mu^2)-{\frac {1}{2}}I_{(0)}({\cal M}^2)+{\frac
{1}{2}}{\cal M}^2 I_{(1)}({\cal M}^2)+{\frac
{m^2}{\beta^2}}-J_r\Phi]\} \nonumber
\\ && \times \exp\{-\delta\int d^2 r {\cal H}_I({\frac {\delta}{\delta J_r}},\Phi,\mu)\}\exp\{{\frac {1}{2}}\int d^2 r_1
     d^2 r_2  J_{r_1}f^{-1}_{r_1 r_2}J_{r_2}\}
\end{eqnarray}
by finishing the Gaussian integral
\begin{eqnarray}
&&\int {\cal D}\phi \exp\{-\int d^2 r [{\frac
{1}{2}}\phi_r(-\nabla^2_r+\mu^2)\phi_r-J_r\phi_r]\}\nonumber \\&=&
  \exp\{-{\frac {1}{2}}\int d^2 r I_{(0)}(\mu^2)\}\exp\{{\frac
  {1}{2}}Jf^{-1}J\}\;.
\end{eqnarray}
Here, $Jf^{-1}J\equiv\int d^2 r_1 d^2 r_2 J_{r_1}f^{-1}_{r_1
r_2}J_{r_2}$ with
\begin{equation}
f^{-1}_{r_1 r_2}=\int {\frac {d^2 p}{(2\pi)^2}} {\frac {1}{p^2+\mu^2}} e^{ip\cdot (r_2-r_1)}={\frac
{1}{2\pi}}K_0(\mu|r_2-r_1|)\;.
\end{equation}
In Eq.(11), $K_0(\mu|r_2-r_1|)$ is the modified Bessel function of
the second kind. Thus, correspondingly to the above modifications,
$W[J]$ becomes $W[J;\Phi,\delta]=\ln(Z[J;\Phi,\delta])$.
Cumulatively expanding $\exp\{-\delta\int d^2 r {\cal H}_I({\frac
{\delta}{\delta J_r}},\Phi,\mu)\}$ and $\ln(\exp\{-\delta\int d^2
r {\cal H}_I({\frac {\delta}{\delta J_r}},\Phi,\mu)\}\exp\{{\frac
{1}{2}}Jf^{-1}J\})$ with Taylor series of the exponential and
logarithmic functions, one has
\begin{eqnarray}
W[J;\Phi,\delta]&=&-\int d^2 r [{\frac {1}{2}} I_{(0)}(\mu^2)-{\frac {1}{2}}I_{(0)}({\cal M}^2)+{\frac
    {1}{2}}{\cal M}^2 I_{(1)}({\cal M}^2)+{\frac {m^2}{\beta^2}}-J_r\Phi]+{\frac {1}{2}}Jf^{-1}J \nonumber\\
      &&+\sum_{l=1}^\infty {\frac {(-1)^{l+1}}{l}}\biggl[\exp\{-{\frac {1}{2}}Jf^{-1}J\} \nonumber \\
      && \ \ \ \ \ \ \ \times \sum_{n=1}^\infty {\frac {(-1)^{n}}{n!}}\delta^n \int \prod_{k=1}^n d^2 r_k {\cal H}_I({\frac
     {\delta}{\delta J_{r_k}}},\Phi,\mu)\exp\{{\frac {1}{2}}Jf^{-1}J\}\biggl]^l \;.
\end{eqnarray}
Substituting Eq.(12) into Eqs.(4) and (5) will give rise to an
expansion of the sG EP which is independent of the parameter $\mu$
for the extrapolating case of $\delta=1$. If the series is
truncated at any order of $\delta$, then the truncated result will
depend on arbitrary parameters $\mu$ and $\Phi$. The background
field $\Phi$ does not affect the EP as long as the wave function
renormalization is not involved and so $\Phi$ can conveniently and
directly be rendered as $\varphi_r$, the vacuum expectation value
of the field operator $\phi$ \cite{13,14}. As for $\mu$, it should
be determined according to the principle of minimal sensitivity
\cite{19}. That is, under the principle of minimal sensitivity,
$\mu$ will be chosen from roots which make the first (or second)
derivative of the truncated result with respect to $\mu$ vanish
\cite{19,13,14}. Thus, $\mu$ will depend on the truncated order.
It is this dependence that makes the truncated result approach the
exact EP order by order. Consequently, the above procedure
provides an approximate method of calculating the sG EP. It should
be noted that in the approximation, taking $\Phi=\phi_r$ in Eq.(4)
will yield $J$ which should be approximated up to the same order
of $\delta$ as the truncated $W[J;\Phi,\delta]$ is.

\section{Approximating sG EP up to the Second Order}
\label{3}

 Now, we concretely calculate the sG EP up to the second order. Using the formulae
$(x+y)^n=\sum_{k=0}^n C_n^k x^{n-k} y^k$, one can have the
following expressions
\begin{eqnarray}
&&\int \prod_{k=1}^n d^2 r_k {\cal H}_I({\frac
     {\delta}{\delta J_{r_k}}},\Phi,\mu)\exp\{{\frac
     {1}{2}}Jf^{-1}J\}\nonumber \\ &=&(-1)^n \int \prod_{k=1}^n d^2 r_k \sum^n_{k=0} C_n^k ({\frac {\mu^2}{2}})^{n-k}
     ({\frac {m^2}{\beta^2}})^k \exp\{{\frac {k \beta^2}{2}}I_{(1)}({\cal
     M}^2)\}\nonumber \\ && \times \prod_{j_1=1}^{n-k}{\frac {\delta^2}{\delta J_{r_{j_1}}^2}}
     \prod_{j_2=n-k+1}^n \cos(\beta({\frac
     {\delta}{\delta J_{r_{j_2}}}}+\Phi))\exp\{{\frac {1}{2}}Jf^{-1}J\}
\end{eqnarray}
and
\begin{eqnarray}
&&\prod_{j=1}^k \cos(\beta({\frac {\delta}{\delta
    J_{r_{j}}}}+\Phi))\exp\{{\frac {1}{2}}Jf^{-1}J\}\nonumber \\ &=&2^{-k}
   \sum^k_{j=0} C_k^j \exp\{i(k-2j)\beta\Phi\}\exp\{{\frac {1}{2}}\int d^2
   r' d^2 r'' [J_{r'}+i\beta(\sum^{k-j}_{j_1=1}\delta(r'-r_{j_1})
   \nonumber \\ && -\sum^{k}_{j_2=k-j+1}\delta(r''-r_{j_2}))]
   f^{-1}_{r' r''}[J_{r''}+i\beta(\sum^{k-j}_{j_1=1}\delta(r''-r_{j_1})-\sum^{k}_{j_2=k-j+1}\delta(r''-r_{j_2}))]
   \}  \;.
\end{eqnarray}
To obtain Eq.(14), we brought its left back to its original
functional integral expression and used the exponential definition
of the cosine function as well as the result of the Gaussian
integral, Eq.(10). From Eq.(14), one can write down
\begin{equation}
\cos(\beta({\frac {\delta}{\delta
    J_{r}}}+\Phi))\exp\{{\frac {1}{2}}Jf^{-1}J\}=\exp\{-{\frac {\beta^2}{2}}f^{-1}_{r
    r}\}\cos(\beta(\int d^2 r'f^{-1}_{r'r}J_{r'}+\Phi))\exp\{{\frac {1}{2}}Jf^{-1}J\}
\end{equation}
and
\begin{eqnarray}
&&\cos(\beta({\frac {\delta}{\delta
    J_{r_{1}}}}+\Phi))\cos(\beta({\frac {\delta}{\delta
    J_{r_{2}}}}+\Phi))\exp\{{\frac {1}{2}}Jf^{-1}J\}\nonumber \\ &=&\exp\{-\beta^2
    (f^{-1}_{r_1 r_1}+f^{-1}_{r_1 r_2})\}\cos(\beta(\int d^2
    r'(f^{-1}_{r'r_1}+f^{-1}_{r'r_2})J_{r'}+2\Phi))\nonumber\\ &&+\exp\{-\beta^2
    (f^{-1}_{r_1 r_1}-f^{-1}_{r_1 r_2})\}\exp\{-i\beta\int d^2
    r'(f^{-1}_{r'r_1}-f^{-1}_{r'r_2})J_{r'}\}\exp\{{\frac {1}{2}}Jf^{-1}J\} \;.
\end{eqnarray}
Employing Eqs.(13), (15) and (16), one can truncate Eq.(12) at the
zeroth, first and second orders of $\delta$ to get
$W^0[J;\Phi,\delta]$, $W^I[J;\Phi,\delta]$ and
$W^{II}[J;\Phi,\delta]$, respectively and then obtain the
approximated EP up to the corresponding orders. (Here, the Greek
number means up to the corresponding order of $\delta$.)

At the zeroth order, $W^0[J;\Phi,\delta]$ is just the first two
terms in Eq.(12) and taking the zeroth order expression of Eq.(4)
as $\Phi$ leads to $J^0_r=0$. Thus, the sG EP at the zeroth order
is
\begin{equation}
{\cal V}^{(0)}(\Phi)={\frac {1}{2}} I_0(\mu^2)
  -{\frac {1}{2}}I_0({\cal
     M}^2) +{\frac {{\cal M}^2}{4}}I_1({\cal M}^2)+ {\frac {m^2}{\beta^2}} \;.
\end{equation}

Up to the first order, one has
\begin{eqnarray}
W^I[J;\Phi,\delta]&=W^0[J;\Phi,\delta]&+\delta\int d^2 r'\{{\frac
{\mu^2}{2}}[f^{-1}_{r'
    r'}+(\int d^2 r''f^{-1}_{r''r'}J_{r''})^2]\nonumber \\ && +{\frac
    {m^2}{\beta^2}}e^{-{\frac {\beta^2}{2}} (f^{-1}_{r' r'}-I_{(1)}[{\cal M}^2])}\cos(\int d^2
    r''f^{-1}_{r''r'}J_{r''}+\beta\Phi)\} \;,
\end{eqnarray}
and  $-{\frac {\delta W^I[J;\Phi,\delta]}{\delta J_r}}= \varphi_r^{I}=\Phi$ yields
\begin{eqnarray}
 &\int d^2 r'f^{-1}_{r'r}J_{r'}+\delta &\int d^2 r'f^{-1}_{r'r}\{\mu^2 \int d^2 r''f^{-1}_{r''r'}J_{r''}\nonumber \\
&&+{\frac {m^2}{\beta^2}}e^{-{\frac {\beta^2}{2}} (f^{-1}_{r'
r'}-I_{(1)}[{\cal M}^2])}\sin(\int d^2
    r''f^{-1}_{r''r'}J_{r''}+\beta\Phi)\}=0 \;.
\end{eqnarray}
From Eq.(19), $J_r$ up to the first order is
\begin{equation}
J^{I}=\delta {\frac {m^2}{\beta^2}}e^{-{\frac {\beta^2}{2}}
(f^{-1}_{r' r'}-I_{(1)}[{\cal M}^2])}\sin(\beta\Phi) \;.
\end{equation}
Because $J^0_r=0$ and there exists no linear term of $J_r$ in
$W^0[J;\Phi,\delta]-\int d^2 r J_r\Phi$ but the quadratic one,
only $J^0_r$ is needed to extract the sG EP up to the first order.
In fact, to obtain the EP up to the $n$th order, one need the
approximated $J$ only up to the $(n-1)$th order. Now one can write
down the sG EP up to the first order
\begin{eqnarray}
{\cal V}^{I}(\Phi,\delta)&=&{\frac
    {m^2}{\beta^2}}+{\frac {1}{2}}(I_{(0)}[\mu^2]-I_{(0)}[{\cal M}^2]) +{\frac {1}{2}} {\cal M}^2 I_{(1)}[{\cal M}^2]
     \nonumber \\  &\ \ \ &  -\delta {\frac
{1}{2}}\mu^2 I_{(1)}[\mu^2] -\delta {\frac
{m^2}{\beta^2}}e^{-{\frac {\beta^2}{2}}
(I_{(1)}[\mu^2]-I_{(1)}[{\cal M}^2])}\cos(\beta\Phi) \;.
\end{eqnarray}

Finally, we consider the second order. When writing down
$W^{II}[J;\Phi,\delta]$, $J_r$ in terms with $\delta^2$ can be
rendered into $J^0_r=0$ (Generally, $J^0_r=0$ can be taken for
terms with the order of $\delta$ at which $W[J;\Phi,\delta]$ is
truncated). Thus,
\begin{eqnarray}
W^{II}[J;\Phi,\delta]&=&W^I[J;\Phi,\delta]+\delta^2 {\frac
{1}{2}}\int d^2 r' d^2 r''\biggl\{{\frac
{1}{2}}\mu^4(f^{-1}_{r'r''})^2\nonumber \\ &&-({\frac
{m^2}{\beta^2}})^2 \exp\{-\beta^2 (f^{-1}_{r'
r'}-I_{(1)}({\cal M}^2))\}\cos^2(\beta\Phi)\nonumber \\
&&-\beta^2\mu^2{\frac {m^2}{\beta^2}} \exp\{-{\frac
{\beta^2}{2}}(f^{-1}_{r' r'}-I_{(1)}({\cal
M}^2))\}(f^{-1}_{r'r''})^2\cos(\beta\Phi)\nonumber\\&& \ \ \ \ \ \
+{\frac {1}{2}}({\frac {m^2}{\beta^2}})^2[\exp\{-\beta^2
    (f^{-1}_{r' r'}-I_{(1)}({\cal M}^2)+f^{-1}_{r' r''})\}\cos(2\beta\Phi)\nonumber \\ &&+\exp\{-\beta^2
    (f^{-1}_{r' r'}-I_{(1)}({\cal M}^2)-f^{-1}_{r' r''})\}]+0(J_r)\biggr\} \;.
\end{eqnarray}
$\varphi_x^{II}=-{\frac {\delta E_0^{II}[J;\Phi,\delta]}{\delta
J_x}}=\Phi$ can be solved for $J^{II}$. In the present case,
however, it is enough to use $J^{I}$. Substituting Eq.(20) into
Eq.(22), keeping terms up to the second order of $\delta$, one
finds that those terms with the factor of squared volume are
cancelled out and the sG EP up to the second order is as follows
\begin{eqnarray}
{\cal V}^{II}&=&{\cal V}^{I}(\Phi,\delta)-\delta^2{\frac
{1}{2}}\{{\frac {1}{2}}\mu^4 I^{(2)}(\mu^2) -{\frac
{m^2}{\beta^2}}\beta^2\mu^2 I^{(2)}(\mu^2)e^{-{\frac {\beta^2}{2}}
(I_{(1)}(\mu^2)-I_{(1)}({\cal M}^2))}\cos(\beta\Phi) \nonumber \\
&\ \ \ & +({\frac {m^2}{\beta^2}})^2\sum^\infty_{k=1}{\frac
{\beta^{2(2k+1)}} {(2k+1)!}}I^{(2k+1)}(\mu^2)
e^{-\beta^2(I_{(1)}(\mu^2)-I_{(1)}({\cal M}^2))}\sin^2(\beta\Phi)
\nonumber \\  &\ \ \ & +({\frac
{m^2}{\beta^2}})^2\sum^\infty_{k=1}{\frac {\beta^{2(2k)}}
{(2k)!}}I^{(2k)}(\mu^2) e^{-\beta^2(I_{(1)}(\mu^2)-I_{(1)}({\cal
M}^2))}\cos^2(\beta\Phi)\}
\end{eqnarray}
with
\begin{eqnarray}
I^{(k)}(\mu^2)&\equiv&{\frac {\int d^2 r' d^2
r''(f^{-1}_{r'r''})^k}{\int d^2 r'
       }}={\frac
{1}{(2\pi)^{k-1}\mu^2}}\int_0^\infty dR R (K_0(R))^k \\ &=&\int
{\frac {d^2 p_1\cdots d^2 p_{k-1} }{(2\pi)^{2(k-1)}}} {\frac
{1}{(p_1^2+\mu^2)\cdots (p_{k-1}^2+\mu^2)
[(p_1+\cdots+p_{k-1})^2+\mu^2]}} \;,
\end{eqnarray}
where we use Eq.(11) to obtain Eq.(24). In Eqs.(21) and (23), the
arbitrary $\mu$ should be determined according to the principle of
minimal sensitivity \cite{13,19}.

In the same way, employing Eqs.(13) and (14), one can obtain the
sG EP up to higher orders, albeit it may be difficult to solve the
renormalization problem and perform numerical calculations. Next
section, we will have a simple discussion on the sG EP up to the
second order, Eq.(23).

For schemes beyond Gaussian approximation, the calculation of
$W[J;\Phi,\delta]$ becomes lengthy and tedious even only up to the
second order. Nevertheless, one can simplify it by using the
replacement trick in Ref.~\cite{12}. Furthermore, if the trick is
combined with a similar procedure to that in our former work
Ref.~\cite{23}, the calculation would become a little more
simplified.

In passing, we point out that one can also obtain the above
results by borrowing Feynman diagrammatic technique \cite{20} with
the propagator of Eq.(11). To make it clear, we rewrite the
interaction Eq.(8) as ${\cal H}_I(\phi_r,\Phi,\mu)=V_q
\phi_r^2+\sum_{n=0}^{\infty}V_{cn}\phi^{2n}+\sum_{n=0}^{\infty}V_{sn}\phi^{2n+1}$.
Here, $V_q=-\mu^2$ is the coefficient attached to the vertex with
two legs, $V_{cn}=-{\frac {m^2}{\beta^2}}\exp\{{\frac
{\beta^2}{2}}I_{(1)}({\cal M}^2)\}\cos(\beta\Phi)(-1)^n {\frac
{\beta^{2n}}{(2n)!}}$ the coefficient attached to vertices with
$2n$ legs and $V_{sn}={\frac {m^2}{\beta^2}}\exp\{{\frac
{\beta^2}{2}}I_{(1)}({\cal M}^2)\}\sin(\beta\Phi)(-1)^n {\frac
{\beta^{2n+1}}{(2n+1)!}}$ those attached to vertices with $(2n+1)$
legs ($n=0,1,2, \cdots$). Obviously, the sG EP is just the sum of
connected one-particle-irreducible vacuum diagrams consisting of
the propagator and $V_q$-,$V_{sn}$- and $V_{cn}$-vertices. For
example, in the great bracket of Eq.(23), the first term comes
from the diagram consisting of two $V_q$-vertices, the second term
is the sum of diagrams consisting of one $V_q$- and one
$V_{cn}$-vertices, the third the sum of diagrams consisting of two
$V_{sn}$-vertices and the fourth the sum of diagrams consisting of
two $V_{cn}$-vertices. When the diagrammatic technique is used,
one has to be careful to count the symmetric factor for every
topologically equivalent diagram correctly.

Additionally, the sG EP beyond the sG Gaussian EP can also be
calculated in the way of Ref.~\cite{14} and the result in
Ref.~\cite{14} is easily used to produce the sG EP up to the
second order which should be identical to Eq.(23).

\section{Discussions on sG post-Gaussian EP}
\label{4}

Noting Eq.(25) and the results ${\frac
{1}{2}}(I_{(0)}[\mu^2]-I_{(0)}[{\cal M}^2]) +{\frac {1}{2}} {\cal
M}^2 I_{(1)}[{\cal M}^2] -{\frac {1}{2}}\mu^2
I_{(1)}[\mu^2]={\frac {\mu^2-{\cal M}^2}{8\pi}}$ and
$I_{(1)}[\mu^2]-I_{(1)}[{\cal M}^2])=-{\frac {1}{4 \pi}}\ln({\frac
{\mu^2}{{\cal M}^2}})$, one can find that there exist no explicit
divergences in Eqs.(21) and (23) for the extrapolating case of
$\delta=1$. This is because we used the Coleman's normal-ordering
prescription at the beginning. Thus, no further renormalization
procedure is needed for the sG EP up to the first order. As for
Eq.(23), we should not draw the same conclusion without further
investigation because there exist series in it. Employing Eq.(24),
one can easily rewrite Eq.(23) as
\begin{eqnarray}
{\cal V}^{II}(\Phi)&=&{\cal V}^{I}(\Phi,\delta)-\delta^2{\frac
{1}{2}}\{{\frac {1}{2}}\mu^4 I^{(2)}(\mu^2) -{\frac
{m^2}{\beta^2}}\beta^2\mu^2 I^{(2)}(\mu^2)e^{-{\frac {\beta^2}{2}}
(I_{(1)}(\mu^2)-I_{(1)}({\cal M}^2))}\cos(\beta\Phi) \nonumber \\
&\ \ \ & + ({\frac {m^2}{\beta^2}})^2 {\frac
{2\pi}{\mu^2}}[A_s(\beta)\sin^2(\beta\Phi)+A_c(\beta)\cos^2(\beta\Phi)]e^{-\beta^2(I_{(1)}(\mu^2)-I_{(1)}({\cal
M}^2))}\}
\end{eqnarray}
with the coefficients $A_s(\beta)=\int_0^\infty dR R [\sinh({\frac
{\beta^2}{2\pi}}K_0(R))-{\frac {\beta^2}{2\pi}}K_0(R)]$ and
$A_c(\beta)=\int_0^\infty dR R [\cosh({\frac
{\beta^2}{2\pi}}K_0(R))-1]$. The Mathematica programm can readily
give values of $A_s(\beta)$ and $A_c(\beta)$ for the case of
$\beta^2\leq 3.431556\pi$ and $\beta^2\leq 3.415399\pi$,
respectively, and one has $A_c(\beta)>A_s(\beta)>0$. For larger
values of $\beta^2$, Mathematica program shows that $A_s(\beta)$
and $A_c(\beta)$ are not convergent. In fact, for the range of
$4\pi\le \beta^2<8\pi$, divergences which can not be removed with
the help of Coleman's normal-ordering prescription were observed
early in 1977 \cite{21}, and it has been shown that they can be
eliminated by the introduction of constant counterterms or in some
alternative ways \cite{22}. Here, we will discuss Eq.(26) only for
the range of $\beta^2\leq 3.4\pi$.

First we give a brief discussion on the sG EP up to the first
order, Eq.(21). Taking $\delta=1$, Eq.(21) is easily rewritten as
\begin{equation}
{\cal V}^{I}(\Phi)= {\frac {m^2}{\beta^2}}+{\frac {\mu^2-{\cal M }^2}{8\pi}}-{\frac {m^2}{\beta^2}}({\frac
{\mu^2}{{\cal M }^2}})^{{\frac {\beta^2}{8 \pi}}} \cos(\beta \Phi) \;.
\end{equation}
To determine $\mu$, one can require ${\frac {d {\cal
V}^{I}(\Phi)}{d\mu}}=0$, which gives rise to $\mu={\cal M }({\frac
{m^2\cos(\beta \Phi)}{{\cal M }^2}})^{{\frac {4\pi}{8
\pi-\beta^2}}}$. Defining the renormalized mass $m_R$ in the same
way as Ref.~\cite{6}, $i.e.$, $m_R^2\equiv {\frac {d^2 {\cal
V}^{I}(\Phi)}{d\Phi^2}}|_{\Phi=0}$, one has $m_R^2={\cal M
}^2({\frac {m^2}{{\cal M }^2}})^{{\frac {8 \pi}{8 \pi-\beta^2}}}$.
If the normal-ordering mass ${\cal M }$ is taken as $m_R$, then
Eq.(27) is just Eq.(2.13) in Ref.~\cite{6} except for a divergent
constant. That is, the sG EP up to the first order is nothing but
the sG Gaussian EP. Thus, the scheme in the present paper provides
a systematic tool of improving the sG Gaussian EP. Note that $\mu$
has no real value when $\cos(\beta \Phi)<0$. In this case, because
the sG Gaussian EP was originally obtained variationally with the
variational parameter $\mu$ \cite{6}, one usually consider ${\cal
V}^{I}(\Phi)$ at the end points of the range $0\leq\mu<\infty$ to
choose the minimal one as the Gaussian EP \cite{6,7}. Thus, the sG
Gaussian EP is a constant for the case of $\cos(\beta \Phi)\leq 0$
because $\mu$ should be chosen as zero, and accordingly, the sG
Gaussian EP is not smooth when $\cos(\beta \Phi)= 0$.

Now we analyze the sG EP up to the second order, Eq.(26). It is
the next order result to the sG Gaussian EP and usually called
post-Gaussian EP \cite{13}. Finishing integrals in Eq.(26) and
taking $\delta=1$, one can reach
\begin{eqnarray}
{\cal V}^{II}(\Phi)&=&{\frac {m^2}{\beta^2}}-{\frac {{\cal M }^2}{8\pi}}+{\frac {\mu^2}{16\pi}}+({\frac
                {m^2}{8\pi}}-{\frac {m^2}{\beta^2}})({\frac {\mu^2}{{\cal M }^2}})^{{\frac {\beta^2}{8 \pi}}}
\cos(\beta \Phi)\nonumber \\ &&-{\frac {\pi}{M^2}}({\frac {m^2}{\beta^2}})^2
  [A_s(\beta)\sin^2(\beta\Phi)+A_c(\beta)\cos^2(\beta\Phi)]({\frac {\mu^2}{{\cal M }^2}})^{{\frac {\beta^2}{4
               \pi}}-1} \;.
\end{eqnarray}
Note that the stationary condition ${\frac {d {\cal
V}^{II}(\Phi)}{d\mu}}=0$ has no real root for $\mu$, and, thus,
one has to appeal to ${\frac {d^2 {\cal
V}^{II}(\Phi)}{d\mu^2}}=0$. It can be explicitly solved with
Mathematica and yields
\begin{eqnarray}
\mu^{{\frac {\beta^2}{8\pi}}-1}&=&\{\beta^2(4\pi-\beta^2)^{{\frac
{1}{2}}}(8\pi-\beta^2)\cos(\beta \Phi)+[\beta^4
(4\pi-\beta^2)(8\pi-\beta^2)^2 \cos^2(\beta\Phi)\nonumber \\ &&\ \
\ \ \ +2048 \pi^4
(6\pi-\beta^2)(A_s(\beta)\sin^2(\beta\Phi)+A_c(\beta)\cos^2(\beta\Phi))]^{{\frac
{1}{2}}} \}\nonumber \\ && \times {\frac { 2^{-5}\pi^{-2}{\cal
M}^2 m^{-2}\beta^2 }{(4\pi-\beta^2)^{{\frac
{1}{2}}}(6\pi-\beta^2)[A_c(\beta)+A_s(\beta)+(A_c(\beta)-A_s(\beta))\cos(2\beta\Phi)]}}\;.
\end{eqnarray}
Obviously, Eq.(29) gives a real $\mu$ for the case of
$\beta^2<4\pi$ (in the case of finite $A_s(\beta)$ and
$A_c(\beta)$), and so the sG post-Gaussian EP can have an explicit
expression by substituting Eq.(29) into Eq.(28). Taking ${\cal
M}=m=1$, one can turn Eqs.(28) and (29) dimensionless for a
numerical calculation. Comparing with the classical potential and
Gaussian EP, the sG post-Gaussian EP possesses the following
features for the case of $\beta^2\leq 3.4\pi$. Firstly, like the
sG Gaussian EP \cite{6,7}, the sG post-Gaussian EP has the same
periodicity as the sG classical potential. This can be seen from
Fig.1 and Fig.2.
\begin{figure}[t]
\includegraphics{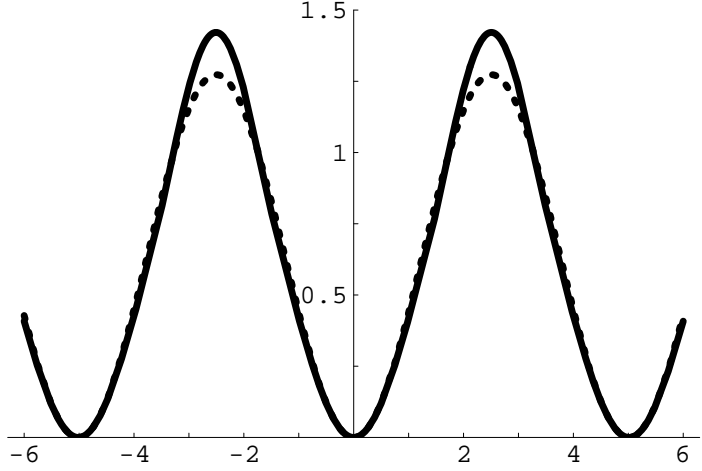}
\caption{\label{fig:1} Comparison between the sG post-Gaussian EP
(solid curve) and classical potential (dotted curve) at
$\beta^2=0.5\pi$. The longitudinal axis represents potentials
$V(\phi)$ or ${\cal V}^{II}(\Phi)$) and the horizontal axis
represents the classical field $\phi$ or $\Phi$.}
\end{figure}
\begin{figure}[h]
\includegraphics{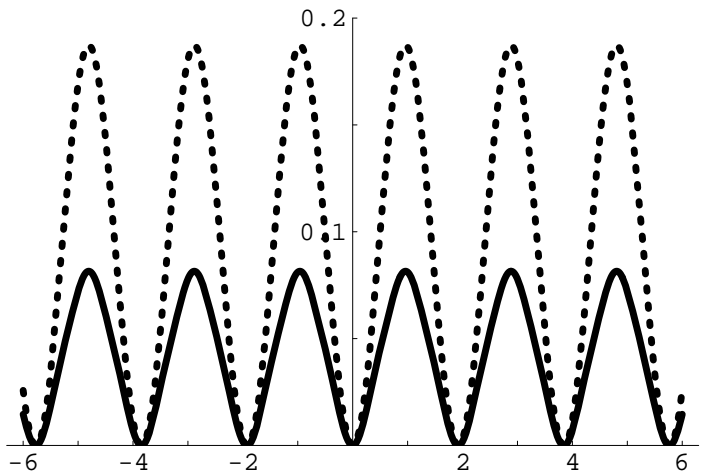}
\caption{\label{fig:2} Similar to Fig.1 but at $\beta^2=3.4\pi$.}
\end{figure}
In Figs.1 and 2, the sG post-Gaussian EPs (solid curves) are
compared with the classical potentials (dotted curves) at
$\beta^2=0.5\pi$ and $3.4\pi$, respectively. In Figs.1 and 2 as
well as the latter Figs.3 and 4, the longitudinal axes represent
potentials $V(\phi), {\cal V}^{I}(\Phi)$ or ${\cal V}^{II}(\Phi)$
and the horizontal axes represent $\phi$ or $\Phi$.
\begin{figure}[t]
\includegraphics{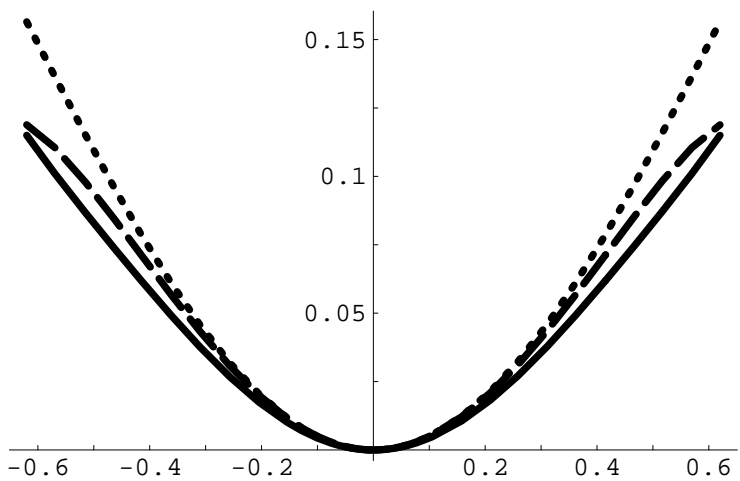}
\caption{\label{fig:3} Comparison between the classical potential
(dotted curve), the sG Gaussian (dashed curve) and post-Gaussian
EP (solid curve) at $\beta^2=2\pi$. The longitudinal axis
represents potentials $V(\phi), {\cal V}^{I}(\Phi)$ or ${\cal
V}^{II}(\Phi)$) and the horizontal axis represents the classical
field $\phi$ or $\Phi$.}
\end{figure}
Secondly, the sG post-Gaussian EP is well defined and smooth for
the whole domain and Figs.1 and 2 illustrate this point. Thirdly,
one observes that in Figs.1 and 2, whereas for smaller $\beta^2$
peaks of the sG post-Gaussian EP are higher than those of the
classical potential, for larger $\beta^2$ (still $\leq 3.4 \pi$)
peaks of the sG post-Gaussian EP are lower than those of the
classical potentials. Finally, Figs.1 and 2 suggest wells of the
sG post-Gaussian EP are wider than those of the classical
potentials. Figs.3 and 4 give more detailed comparison on the last
point among wells of the classical potentials (dotted), the sG
Gaussian (dashed) and the post-Gaussian EPs (solid) for
$\beta^2=2\pi$ and $3.4\pi$, respectively.
\begin{figure}[h]
\includegraphics{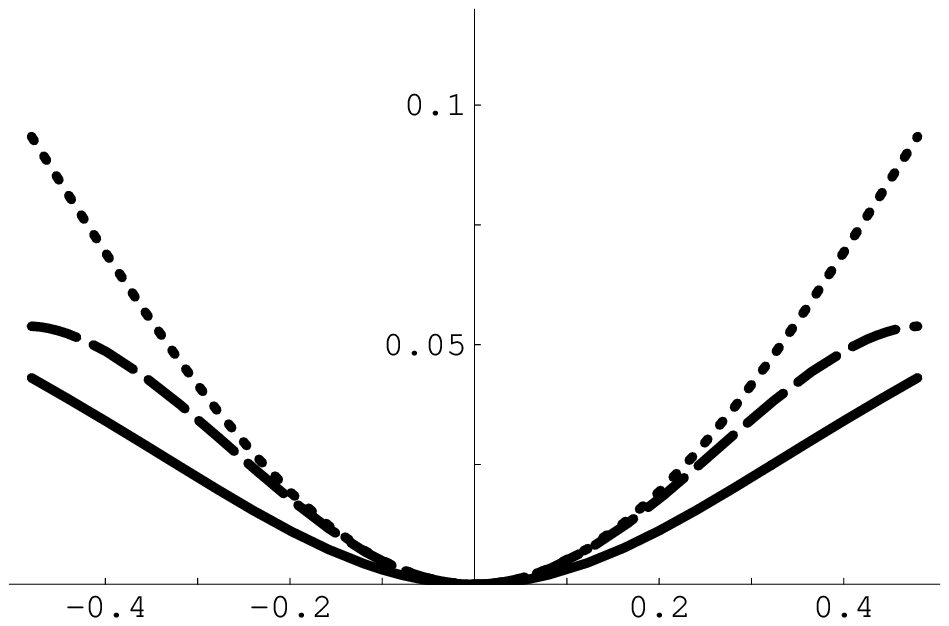}
\caption{\label{fig:4} Similar to Fig.3 but at $\beta^2=3.4\pi$.}
\end{figure}
 Noting that in Fig.1, the wells of the sG post-Gaussian EP are almost
identical to those of the classical potentials, one can conclude
that the correction of the sG post-Gaussian EP from the sG
Gaussian EP becomes more substantial with the increase of
$\beta^2$.

\section{Conclusion}
\label{5}

In this paper, we have performed an optimized expansion in the
spirit of the background field method plus the Coleman's
normal-ordering renormalization prescription to calculate the EP
of sG field theory beyond the Gaussian approximation. We obtained
the sG EP up to the second order, Eq.(28) together with Eq.(29),
which is finite for the range of $\beta\leq 3.4\pi$ (an
approximate value from Mathematica) without the need of any
further renormalization procedure. It is well-defined for the
whole domain of $\Phi$, and give more substantial correction to
the sG Gaussian EP with the increase of $\beta^2$. In view of the
existence of many equivalents to the sG model and the importance
of going beyond the Gaussian approximation, we believe that our
investigation in this paper is meaningful and interesting. Still,
some further investigations are needed. Without any doubt, it will
be important to renormalize Eq.(23) beyond the range of $\beta^2$
which we have treated in the present paper and apply the result to
many sister systems of the sG model. Additionally, a
generalization of the present work to a finite-temperature case
will be also interesting because the finite-temperature effect at
the 1L and Gaussian approximation may change the periodicity of
the classical potential \cite{1,2,3,4,7}.

\begin{acknowledgments}
 Lu acknowledges Mr. J.-G. Bu, Prof. J. H. Yee and Prof. Dr. H. Kleinert for their useful discussions. This project was
 supported by Korea Research  Foundation (2001-005-D00011) and also by the Korea Science and Engineering Foundation
through the Center for Strongly correlated materials Research
(SNU). Lu's work was also supported in part by the National
Natural Science Foundation of China under the grant No. 19875034.
\end{acknowledgments}


\begin{thebibliography}{99}
\bibitem{1} G. P. Malik, J. Subba Rao and A. Goyal, Phys. Rev. D {\bf 21}
            (1980) 2421.
\bibitem{2} K. Babu Joseph and V. C. Kuriakose, Phys. Lett. A {\bf 88} (1982)
            447.
\bibitem{3} W. H. Kye, S. I. Hong and J. K. Kim, Phys. Rev. D {\bf 45} (1992)
            3006.
\bibitem{4} D. K. Kim, S. I. Hong, M. H. Lee and I. G. Koh, Phys. Lett. A
            (1994) 379.
\bibitem{5} K. Yoshida and W. Souma, Phys. Rev. D {\bf 64} (2001) 125002.
\bibitem{6} R. Ingermanson, Nucl. Phys. B {\bf 266} (1986) 620.
\bibitem{7} P. Roy, R. Roychoudhury and Y. P. Varshni, Mod. Phys. Lett. A
            {\bf 4} (1989) 2031.
\bibitem{8} B. W. Xu and Y. M. Zhang, J. Phys. A {\bf 25} (1992) L1039;
            Y. M. Zhang, B. W. Xu and W. F. Lu, Phys. Rev. B {\bf 49} (1994) 854.
\bibitem{9} G. J. Ni, S. Y. Lou, S. Q. Chen and H. C. Lee,
            Phys. Rev. B {\bf 41} (1990) 6947; Z. J. Chen, Y. M Zhang and B. W. Xu, {\it ibid.} B {\bf 49} (1994)
            12535; B. W. Xu and Y. M. Zhang, Phys. Rev. B {\bf 50} (1994) 18651.
\bibitem{10}W. F. Lu, B. W. Xu and Y. M. Zhang, Phys. Lett. B {\bf 309}
            (1993) 109.
\bibitem{11}E. K. Sklyanin, L. A. Takhtadzyan and L. D. Faddeev, Theor. Math. Phys. {\bf 40} (1979) 688;
            S. Lukyanov and A. Zamolodchikov, Nucl. Phys. B {\bf 493} (1997) 571; A. O. Gogolin, A. A. Nersesyan and A. M.
            Tsvelik, {\it Bosonization and Strongly Correlated Systems}, Cambridge University Press, Cambridge, England,
            1998, Chapter 10.
\bibitem{12}To see Chapter 19 in the book: H. Kleinert and V. Schulte-Frohlinde, {\it Critical
            Properties of $\phi^4$-Theories}, World Scientific, Singapore, 2001.
\bibitem{13}I. Stancu and P. M. Stenvenson,Phys. Rev. D {\bf 42} (1990) 2710; I. Stancu, {\it ibid.} D {\bf 43} (1991) 1283.
\bibitem{14}W. F. Lu, C. K. Kim and K. Nham, Phys. Lett. B {\bf 540} (2002) 309, or see hep-th/0204046.
\bibitem{15}S. Coleman, Phys. Rev. D {\bf 11} (1975) 2088;
            S. J. Chang, Phys. Rev. D {\bf 13} (1976) 2778.
\bibitem{16}W. F. Lu and C. K. Kim, J. Phys. A {\bf 35} (2002) 393.
\bibitem{17}P. Ramond, {\it Field Theory: a Modern Primer}, Revised
            Printing, Addison-Wesley, New York, 1990, \S 3.7.
\bibitem{18}C. Itzykson and J.-B. Zuber, {\it Quantum Field
            Theory},McGraw-Hill Inc., New York, 1980, \S 6-2-4.
\bibitem{19}P. M. Stevenson, Phys. Rev. D {\bf 23} (1981) 2916;
            Phys. Lett. B {\bf 100} (1981) 61; Phys. Rev. D {\bf 24} (1981)
            1622; S. K. Kaufmann and S. M. Perez, J. Phys. A {\bf 17} (1984)
            2027; P. M. Stevenson, Nucl. Phys. B {\bf 231} (1984) 65.
\bibitem{20}S. Samuel, Phys. Rev. D {\bf 18} (1978) 1916; D. J. Admit, Y. Y. Goldschmidt, and G. Grindstein,
            J. Phys. A {\bf 13} (1980) 585.
\bibitem{21}J. A. Swieca, Fortschr. Phys. {\bf 25} (1977) 303; B. Schroer and T. Truong, Phys. Rev. D {\bf 15} (1977) 1684.
\bibitem{22}G. Gallavotti, Rev. Mod. Phys. {\bf 57} (1985) 471; A. Lima-Santos and E.C. Marino, J. Stat. Phys. {\bf 55}
            (1989) 157.

\bibitem{23} W. F. Lu, S. K. You, J. Bak,
            C. K. Kim and K. Nahm, J. Phys. A {\bf 35} (2002) 21.
\end{thebibliography}
\end{document}